\documentclass[letterpaper,twocolumn,,showpacs,prb
superscriptaddress,email]{revtex4}

\usepackage[pdftex]{graphicx}  
\usepackage{amssymb}
\usepackage{amsmath,amsfonts,latexsym}
\usepackage{array,tabularx,color}
\usepackage{dcolumn}               
\usepackage[normalem]{ulem}
\usepackage{comment}
\usepackage{bm}
\usepackage[latin1]{inputenc}
\usepackage{color}

\DeclareMathOperator*{\Simiq}{\simeq}

\newcommand{\vect}[1]{{\mathbf #1}}

\newcommand{\Frac}[2]{\displaystyle\frac{#1}{#2}}

\begin{document}

\title{Drag in a resonantly driven polariton fluid}

\author{A.~C.~Berceanu}
\email[Corresponding author: ]{andrei.berceanu@uam.es}
\affiliation{Departamento de F\'isica Te\'orica de la Materia
  Condensada, Universidad Aut\'onoma de Madrid, Madrid 28049, Spain}

\author{E.~Cancellieri\footnote{Present address:
    Laboratoire Kastler Brossel, Universit\'e Pierre et Marie Curie,
    \'Ecole Normale Superieure, CNRS, 4 place Jussieu, 75005 Paris,
    France}}
\affiliation{Departamento de F\'isica Te\'orica de la Materia
  Condensada, Universidad Aut\'onoma de Madrid, Madrid 28049, Spain}

\author{F. M. Marchetti}
\affiliation{Departamento de F\'isica Te\'orica de la Materia
  Condensada, Universidad Aut\'onoma de Madrid, Madrid 28049, Spain}

\date{\today}

\begin{abstract}
  We study the linear response of a coherently driven polariton fluid
  in the pump-only configuration scattering against a point-like
  defect and evaluate analytically the drag force exerted by the fluid
  on the defect. When the system is excited near the bottom of the
  lower polariton dispersion, the sign of the interaction-renormalised
  pump detuning classifies the collective excitation spectra in three
  different categories~[C. Ciuti and I. Carusotto, \emph{physica
      status solidi (b)} \textbf{242}, 2224 (2005)]: linear for zero,
  diffusive-like for positive, and gapped for negative detuning. We
  show that both cases of zero and positive detuning share a
  qualitatively similar crossover of the drag force from the subsonic
  to the supersonic regime as a function of the fluid velocity, with a
  critical velocity given by the speed of sound found for the linear
  regime. In contrast, for gapped spectra, we find that the critical
  velocity exceeds the speed of sound. In all cases, the residual drag
  force in the subcritical regime depends on the polariton lifetime
  only. Also, well below the critical velocity, the drag force varies
  linearly with the polariton lifetime, in agreement with previous
  work~[E. Cancellieri \emph{et al.}, \emph{Phys. Rev. B} \textbf{82},
    224512 (2010)], where the drag was determined numerically for a
  finite-size defect.
\end{abstract}





\maketitle

\section{Introduction}
\label{sec:intro}
Out of equilibrium quantum fluids such as polaritons in semiconductor
microcavities are being the subject of an intensive study. Microcavity
polaritons, the quasiparticles resulting from the strong coupling of
cavity photons and quantum well excitons, have the prerogative of
being easy to both manipulate, via an external laser, and detect, via
the light escaping from the cavity~\cite{kavokin_laussy}. In
particular, resonant excitation allows the accurate tuning of the
fluid properties, such as its density and current. However, the
polariton lifetime being finite establishes the system as
intrinsically out of equilibrium: An external pump is needed to
continuously replenish the cavity of polaritons, that quickly, on a
scale of tens of picoseconds, escape.

Recently, the superfluid properties of a resonantly pumped polariton
quantum fluid in the pump-only configuration --- i.e., where no other
states aside the pump one are occupied by, e.g., parametric scattering
--- have been actively investigated both experimentally and
theoretically~\cite{carusotto04,ciuti_05,amo09_b,cancellieri_10,pigeon_11,amo_science11,nardin11,sanvitto11}. This
pumping scheme, differently from other cases, such as the resonant
optical parametric oscillator regime and the non-resonant pumping
scheme, creates a polariton fluid that, inside the pump spot, is not
characterised by a free phase. On the contrary, the phase of the pump
state is locked to the one of the external pumping
laser. Nevertheless, it has been predicted~\cite{carusotto04,ciuti_05}
and observed~\cite{amo09_b} that scattering can be suppressed below a
critical velocity, where the system displays superfluid behaviour,
similarly to what has been predicted by the Landau criterion for
equilibrium superfluid condensates. Further, a fixed phase clearly
prevents the formation of phase dislocations, such as vortices and
solitons. For this reason, it has been suggested~\cite{pigeon_11} and
experimentally realised~\cite{amo_science11} that the defect can be
located just outside the pump spot, where the hydrodynamic nucleation
of vortices, vortex-antivortex pairs, arrays of vortices, and solitons
can be observed when the fluid collides with the extended
defect. Similarly, nucleation of vortices in the wake of the obstacle
has been observed in pulsed experiments ~\cite{nardin11,sanvitto11}.

In a conservative quantum liquid flowing past a small defect, the
Landau criterion for superfluidity links the onset of dissipation at a
critical fluid velocity with the shape of the fluid collective
excitation spectrum~\cite{pitaevskii03}. In particular, for weakly
interacting Bose gases, the dispersion of the low-energy excitation
modes being linear implies that the critical velocity for superflow
coincides with the speed of sound $c_s$. Clearly, this is strictly
correct only for vanishingly small
perturbations~\cite{astrakharchik04}, while for a defect with finite
size and strength, the critical velocity can be smaller than
$c_s$~\cite{onofrio00,iasenelli06}.

However, even for perturbatively weak defects, in out-of-equilibrium
systems, where the spectrum of excitations is complex, the validity of
the Landau criterion has to be
questioned~\cite{szymanska06:prl,wouters10_b,cancellieri_10}. In the
particular case of coherently driven polaritons in the pump-only
configuration, it has been predicted~\cite{carusotto04,ciuti_05}, and
later observed~\cite{amo09_b}, that scattering is suppressed at either
strong enough pump powers or small enough flow velocities. Yet, on a
closer scrutiny, it has been shown that, despite the apparent validity
of the Landau criterion, the system always experiences a residual drag
force even in the limit of asymptotically large
densities~\cite{cancellieri_10} or small velocities. This result has
been proven by numerically solving the Gross-Pitaevskii equation
describing the resonantly-driven polariton system in presence of a
non-perturbative extended defect. Here, the drag force exerted by the
defect on the fluid has been shown to display a smooth crossover from
the subsonic to the supersonic regime, similarly to what it has been
found in the case of non-resonantly pumped
polaritons~\cite{wouters10_b}. In this work, we find an even richer
phenomenology for the dependence of the drag force on the fluid
velocity and two different kinds of crossovers from the sub- to the
supercritical regime. Further, we show that the origin of the residual
drag force, which, in agreement with Ref.~\cite{cancellieri_10}, lies
in the polariton lifetime only, can be demonstrated even within a
linear response approximation.

More specifically, in this work, we apply the linear response theory
to analytically evaluate the drag force exerted by the coherently
driven polariton fluid in the pump-only configuration on a point-like
defect. To simplify the formalism, we restrict our analysis to the
case of resonant pumping close to the bottom of the lower polariton
dispersion, where the dispersion is quadratic. Here, the properties of
the collective excitation spectrum have been shown to be uniquely
determined by three parameters only~\cite{ciuti_05}: the fluid
velocity $v_p$, the interaction-renormalised pump detuning $\Delta_p$,
and the polariton lifetime $\kappa$. In particular, the sign of the
detuning $\Delta_p$ determines three qualitatively different types of
spectra: linear for $\Delta_p= 0$, diffusive-like for $\Delta_p> 0$,
and gapped for $\Delta_p< 0$. 

For both cases of linear and diffusive spectra, we find a
qualitatively similar behaviour of the drag force as a function of the
fluid velocity $v_p$: In particular, the drag displays a crossover
from a subsonic or superfluid regime --- characterised by the absence
of quasiparticle excitations --- to a supersonic regime --- where
Cherenkov-like waves are generated by the defect and propagate into
the fluid. The crossover becomes sharper for increasing polariton
lifetimes $1/\kappa$ and displays the typical threshold behaviour for
$\kappa \to 0$ with a critical velocity given by the speed of sound of
linear regime, $v^c= c_s$, exactly as for weakly interacting
equilibrium superfluids (in the case of perturbatively weak
defects). This behaviour is similar to the one predicted for polariton
superfluids excited non-resonantly~\cite{wouters10_b}, where the
spectrum in that case is diffusive-like.

However, for gapped spectra at $\Delta_p <0$, we find that the
critical velocity governing the drag crossover exceeds the speed of
sound, $v^c > c_s$, and we determine an analytical expression of $v^c$
as a function of the detuning $\Delta_p$. Further, for $\kappa \to 0$,
the drag has a threshold-like behaviour qualitatively different from
the one of weakly interacting equilibrium superfluids, with the drag
jumping discontinuously from zero to a finite value at $v_p=v^c$.

We evaluate the drag as a function of the polariton lifetime $\kappa$
and find for all three cases that: In the supercritical regime,
$v_p>v^c$, the lifetime tends to suppress the propagation of the
Cherenkov waves away from the defect and therefore to suppress the
drag. Instead, well in the subcritical regime, $v_p \ll v^c$, we find
that the residual drag goes linearly to zero with the polariton
lifetime $\kappa$, in agreement to what it was found in
Ref.~\cite{cancellieri_10}, by making use of a non-perturbative
numerical analysis for a finite size defect. Similarly to
Ref.~\cite{cancellieri_10}, here, we do also find that the residual
drag in the subcritical regime can be explained in terms of an
asymmetric perturbation induced in the fluid by the defect in the
direction of the fluid velocity.

This paper is structured as follows: In Sec.~\ref{sec:linea} we
briefly introduce the linear response approximation. We classify the
three types of collective excitation spectra in the simplified case of
excitation close to the bottom of the lower polariton dispersion in
Sec.~\ref{sec:spect}. In Sec.~\ref{sec:drag} we derive the drag force
and characterise the crossover from the subsonic to the supersonic
regime in the three cases of zero, positive and negative detuning. In
this section, we also evaluate the drag as a function of the polariton
lifetime, interpreting therefore the results of
Ref.~\cite{cancellieri_10}.  Brief conclusions are drawn is
Sec.~\ref{sec:concl}.

\section{Linear response}
\label{sec:linea}
The description of cavity polaritons resonantly excited by an external
laser is usually formulated in terms of a classical non-linear
Schr\"odinger equation (or Gross-Pitaevskii equation)~\cite{ciuti03}
for the lower polariton (LP) field $\psi_{LP}(\vect{r}, t)$ ($\hbar =
1$):
\begin{multline}
  i \partial_t \psi_{LP} = [\omega_{LP}(-i\nabla) - i\kappa +
    V(\vect{r}) + g |\psi_{LP}|^2]\psi_{LP} \\
  + \mathcal{F}(\vect{r},t)\; .
\label{eq:basic}
\end{multline}
The LP dispersion is expressed in terms of the photon
$\omega_C(\vect{k}) = \omega_C^0 + \frac{\vect{k}^2}{2m_C}$ and
exciton $\omega_X^0$ energies, the photon mass $m_C$, and the Rabi
splitting $\Omega_R$~\cite{kavokin_laussy}:
\begin{multline}
  \omega_{LP}(\vect{k}) = \frac{1}{2} \left[\omega_C(\vect{k}) +
    \omega_X^0\right]\\
  - \frac{1}{2} \sqrt{\left[\omega_C(\vect{k}) -
    \omega_X^0 \right]^2 + \Omega_R^2} \; .
\label{eq:dispe}
\end{multline}
Because polaritons continuously decay at a rate $\kappa$, the cavity
is replenished by a continuous wave resonant pump $F(\vect{r},t)$ at a
wavevector $\vect{k}_p$ (we will later assume $\vect{k}_p$ directed
along the $x$-direction, $\vect{k}_p = (k_p,0)$) and frequency
$\omega_p$:
\begin{equation}
  \mathcal{F}(\vect{r},t) = f_p e^{i (\vect{k}_p \cdot \vect{r} -
    \omega_p t)} \; .
\end{equation}
Note that, as discussed in appendix~\ref{app:full},
Eq.~\eqref{eq:basic} is a simplified description of the polariton
system: This implies that the interaction non-linearities are small
enough not to mix the lower and upper polariton branches. Moreover,
starting from a formulation in terms of coupled exciton and photon
fields, the polariton lifetime would be momentum dependent and,
similarly, the polariton-polariton interaction strength $g$ is not
contact-like as instead assumed in Eq.~\eqref{eq:basic}. However, as
shown in appendix~\ref{app:full}, these simplifications, do not affect
our results qualitatively, rather, allow to write them in terms of
simpler expressions. Further, we have checked that, whenever the
system is excited near the bottom of the lower polariton dispersion,
the results for the drag force reported in Sec.~\ref{sec:drag}
coincide with the ones obtained by using an exact photon-exciton
coupled field description.

The potential $V(\vect{r})$ in Eq.~\eqref{eq:dispe} describes a
defect, which can be either naturally present in the cavity
mirror~\cite{amo09_b} or it can be created by an additional
laser~\cite{amo10}. Later on, we will assume the defect to be
point-like $V(\vect{r})=g_V \delta(\vect{r})$ and weak, so that we can
apply the linear response approximation~\cite{astrakharchik04}.
In this treatment, one divides the response of the LP field in a
mean-field component $\psi_0$ corresponding to the case when the
perturbing potential is absent, and a fluctuation part $\delta \psi
(\vect{r},t)$ reflecting the linear response of the system to the
perturbing potential:
\begin{equation}
  \psi_{LP} (\vect{r},t) = e^{-i \omega_p t} \left[e^{i \vect{k}_p
      \cdot \vect{r}} \psi_0 + \delta \psi (\vect{r},t)\right] \; .
\label{eq:mfield}
\end{equation}
By substituting~\eqref{eq:mfield} into~\eqref{eq:basic}, we obtain a
mean-field equation and, by retaining only the linear terms in the
fluctuation field and the defect potential, the following first order
equation in $\delta \psi (\vect{r},t)$:
\begin{equation}
  i \partial_t \begin{pmatrix} \delta \psi \\ \delta
    \psi^* \end{pmatrix} = \hat{\mathcal{L}} \begin{pmatrix} \delta
    \psi \\ \delta \psi^* \end{pmatrix} + V(\vect{r}) \begin{pmatrix}
    \psi_0 e^{i \vect{k}_p \cdot \vect{r}} \\ -\psi_0^{\star} e^{-i
      \vect{k}_p \cdot \vect{r}}\; ,
    \end{pmatrix}
\label{eq:linre}
\end{equation}
where the operator $\hat{\mathcal{L}}$ is given by:
\begin{equation}
 \hat{\mathcal{L}} = \begin{pmatrix} \widetilde{\omega_{LP}}
   (-i\nabla) - i \kappa & g \psi_0^2 e^{2 i \vect{k}_p \cdot
     \vect{r}} \\ -g {\psi_0^{\star}}^2 e^{-2 i \vect{k}_p \cdot
     \vect{r}}& - \widetilde{\omega_{LP}}(-i \nabla) -
   i\kappa \end{pmatrix}\; ,
\end{equation}
with $\widetilde{\omega_{LP}} = \omega_{LP}-\omega_p + 2g
|\psi_0|^2$. We are not interested here in solving the complex cubic
mean-field equation for $\psi_0$, as this has been already widely
studied~\cite{kavokin_laussy}. Rather, we want to study the response
of the system to the presence of the defect and how different
behaviours of the onset of dissipation can be described in terms of
the different excitation spectra one can get for polaritons resonantly
pumped close to the bottom of the LP dispersion.

\begin{figure}
\begin{center}
\includegraphics[width=1.0\columnwidth,angle=0]{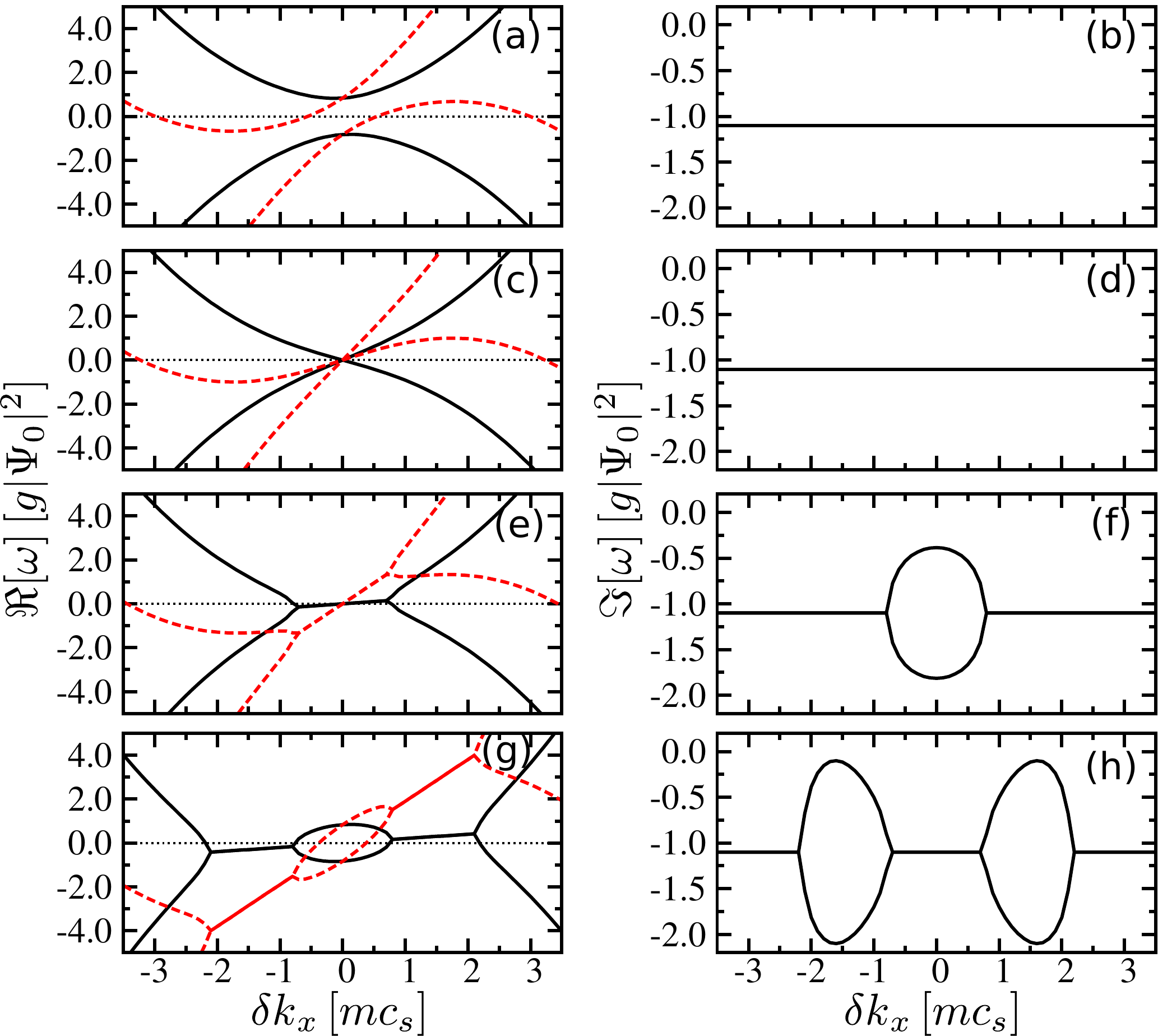}
\end{center}
\caption{(Color online) Collective excitation spectra for the subsonic
  (thick solid [black] line at $v_p=0.2 c_s$, with
  $c_s=\sqrt{g|\psi_0|^2/m}$) and supersonic (dashed [red] line at
  $v_p=1.9 c_s$) regimes and for an interaction-renormalised pump
  detuning $\Delta_p=-0.3 g|\psi_0|^2$ (a, b), $\Delta_p = 0$ (c, d),
  $\Delta_p=0.3g|\psi_0|^2$ (e, f) and $\Delta_p=2.3g|\psi_0|^2$ (g,
  h). Real parts of the spectra are plotted in the left panels and the
  corresponding imaginary parts in the right panels for $\kappa=1.1
  g|\psi_0|^2$ --- note that in our description the spectrum imaginary
  parts do not depend on the fluid velocity $v_p$.}
\label{fig:spect}
\end{figure}
%
\subsection{Spectrum of collective excitations}
\label{sec:spect}
The spectrum of the collective excitations can be obtained by
diagonalising the operator $\hat{\mathcal{L}}$ in the momentum space
representation:
\begin{equation}
  \mathcal{L}_{\vect{k},\vect{k}_p} = \begin{pmatrix}
    \widetilde{\omega_{LP}} (\delta \vect{k}+\vect{k}_p) - i \kappa &
    g \psi_0^2 \\ -g {\psi_0^{\star}}^2 & -
    \widetilde{\omega_{LP}}(\delta \vect{k}-\vect{k}_p) -
    i\kappa \end{pmatrix}\; ,
\label{eq:opell}
\end{equation}
where, $\delta \vect{k} = \vect{k} - \vect{k}_p$. The description of
the spectrum simplifies in the case when the pumping is close to the
bottom of the LP dispersion, that can be approximated as parabolic
\begin{equation}
  \omega_{LP} (\delta \vect{k} \pm \vect{k}_p) \simeq \omega_{LP}(0) +
  \frac{k_p^2}{2m} + \frac{\delta \vect{k}^2}{2m} \pm \delta \vect{k}
  \cdot \vect{v}_p \; ,
\end{equation}
where $\vect{v}_p=\vect{k}_p/m$ is the fluid velocity, and $m$ is the
LP mass, $m = 2m_C [1 - (\omega_C^0 - \omega_X^0)/\sqrt{(\omega_C^0 -
    \omega_X^0)^2 + \Omega_R^2}]^{-1}$. This simplification allows one
to describe the complex spectrum in terms of three parameters only,
namely the fluid velocity $\vect{v}_p$, the interaction-renormalised
pump detuning
\begin{equation}
  \Delta_p = \omega_p - \left[\omega_{LP} (0) +\frac{k_p^2}{2m} +
    g|\psi_0|^2\right]
\end{equation}
and the LP lifetime $\kappa$:
\begin{equation}
  \omega^{\pm} (\vect{k}) = \delta \vect{k}\cdot \vect{v}_p - i\kappa
  \pm \sqrt{\varepsilon(\delta \vect{k}) \left[\varepsilon(\delta
      \vect{k}) + 2g|\psi_0|^2\right]} \; ,
\label{eq:spect}
\end{equation}
where $\varepsilon(\vect{k}) = \frac{k^2}{2m} - \Delta_p$. If energies
are measured in units of the mean-field energy blue-shift $g
|\psi_0|^2$ (we will use the notation $\Delta_p' =
\Delta_p/g|\psi_0|^2$ and $\kappa'= \kappa/g|\psi_0|^2$), then the
fluid velocity $v_p$ is measured in units of the speed of sound $c_s =
\sqrt{g|\psi_0|^2/m}$. In order to make connection with the current
experiments, note that, for blue-shifts in the range $g |\psi_0|^2
\simeq 0.1-1$~meV, typical values of the speed of sound $c_s$ are
$0.8-2.7\times 10^6$~m/s. Similarly, for common values of the LP mass,
the range in momenta in Fig.~\ref{fig:spect} comes of the order of
$\delta k_x \simeq 0.2-0.8$~$\mu$m${}^{-1}$.


The spectrum~\eqref{eq:spect} can be classified according to the sign
of the interaction-renormalised pump detuning
$\Delta_p$~\cite{carusotto04,ciuti_05} --- see
Fig.~\ref{fig:spect}. For $\Delta_p<0$ [panels (a,b)], the real part
of the spectrum is \emph{gapped} while the imaginary part is
determined by the polariton lifetime $\kappa$ only. If one applies the
Landau criterion making reference to the real part of the spectrum
only, then one finds a critical velocity
\begin{equation}
  \Frac{v^c}{c_s} = \sqrt{1 + |\Delta_p'| +
    \sqrt{|\Delta_p'|(|\Delta_p'| + 2)}} > 1\; ,
\label{eq:criti}
\end{equation}
always larger than the speed of sound for $\Delta_p<0$. If the fluid
velocity is subcritical, $v_p<v^c$ (see [black] solid lines in
Fig.~\ref{fig:spect}(a)), then no quasiparticles can be excited and
thus, for infinitely living polaritons $\kappa \to 0$, the fluid would
experience no drag when scattering against the defect. For
supercritical velocities instead, $v_p>v^c$ see [red] dashed lines in
Fig.~\ref{fig:spect}(a), one expects dissipation in the form of
radiation of Cherenkov-like waves from the defect into the fluid. In
the supercritical regime, the set of wavevectors $\vect{k}$ for which
$\Re[\omega^{+} (\vect{k})] = 0$ form a closed curve in the
$\vect{k}$-space with no singularity of the derivative, i.e., in other
words, the radiation can be emitted in all possible directions around
the defect. This, as we will see in the next section, will imply that
the drag force for $\kappa \to 0$ goes abruptly, rather than
continuously, from zero at $v_p<v^c$ to a finite value at $v_p \ge
v^c$.

The spectrum gap closes to zero in the resonant situation at
$\Delta_p=0$, when the two branches $\omega^{\pm} (\vect{k})$ touch at
$\delta \vect{k}=0$ [panels (c,d) of Fig.~\ref{fig:spect}]: Here, the
real part of the spectrum displays the standard \emph{linear
  dispersion} at small wavevectors as for the weakly interacting
bosonic gases, with the slope given by $c_s \pm v_p$. The imaginary
part, as in the previous case, is constant and equal to $-\kappa$. It
is clear therefore that in this case, when $\kappa \to 0$, one
recovers the equilibrium results valid for weakly interacting
gases~\cite{astrakharchik04,carusotto06_prl}, where the critical
velocity for superfluidity equals the speed of sound, $v^c=c_s$, and
the drag displays a threshold like behaviour. Here, in the supersonic
regime $v_p> v^c$, the close curve $\Re[ \omega^{+} (\vect{k})] = 0$
has instead a singularity, resulting in the standard Mach cone of
aperture $\theta$, $\sin \theta = c_s/v_p$, inside which radiation
from the defect cannot be emitted~\cite{carusotto06_prl}.

Finally, for $\Delta_p>0$, the real parts of the particle $\omega^+
(\vect{k})$ and hole $\omega^- (\vect{k})$ branches of the spectrum
touch together in either one [$\Delta_p \le 2$, see panels (e,f)] or
two [$\Delta_p > 2$, see panels (g,h)] separate regions in momentum
space. In the same regions, the corresponding imaginary parts instead
split. With a somewhat abuse of language, we call these kinds of
spectrum, \emph{diffusive-like}. We note that, clearly, these spectra
have no correspondence in equilibrium systems, because a finite
polariton lifetime $\kappa$ is needed in order for these modes to be
stable, $\Im [ \omega^{\pm} (\vect{k}) ]<0$. We also note that for
these spectra, even if considering only the real part of the
collective excitation spectrum, as soon as the fluid is in motion
$v_p>0$, dissipation in the form of waves is possible. However, we
will see that similarly to the case of polaritons non-resonantly
pumped~\cite{wouters10_b}, when decreasing $\kappa$ (and accordingly
$\Delta_p$ in order to have stable solutions), this situation connects
continuously to the previous case, where a threshold-like behaviour
with $v^c = c_s$ was found.

We will see in the next section how these different spectra imply only
two qualitatively different types of crossover of the drag force as a
function of the fluid velocity, for either $\Delta_p<0$ or $\Delta_p
\ge 0$ pump detunings.

\begin{figure}
\begin{center}
\includegraphics[width=1.0\columnwidth,angle=0]{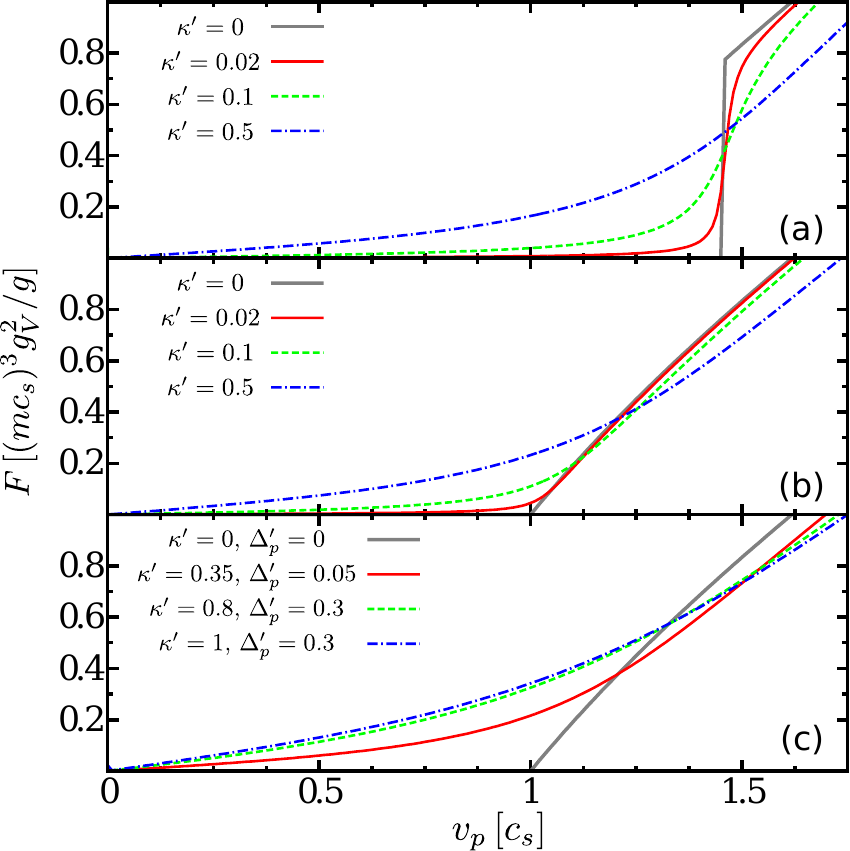}
\end{center}
\caption{(Color online) Drag force $F$ as a function of the fluid
  velocity $v_p$ for different values of the pump detuning $\Delta_p$:
  $\Delta_p=-0.3g|\Psi_0|^2$ (a), $\Delta_p=0$ (b), and $\Delta_p>0$
  (c), and for different values of the polariton lifetime --- here, we
  use the notation $\kappa' = \kappa/g|\psi_0|^2$, $\Delta_p' =
  \Delta/g|\psi_0|^2$.}
\label{fig:dragv}
\end{figure}
\begin{figure}
\begin{center}
\includegraphics[width=1.0\columnwidth,angle=0]{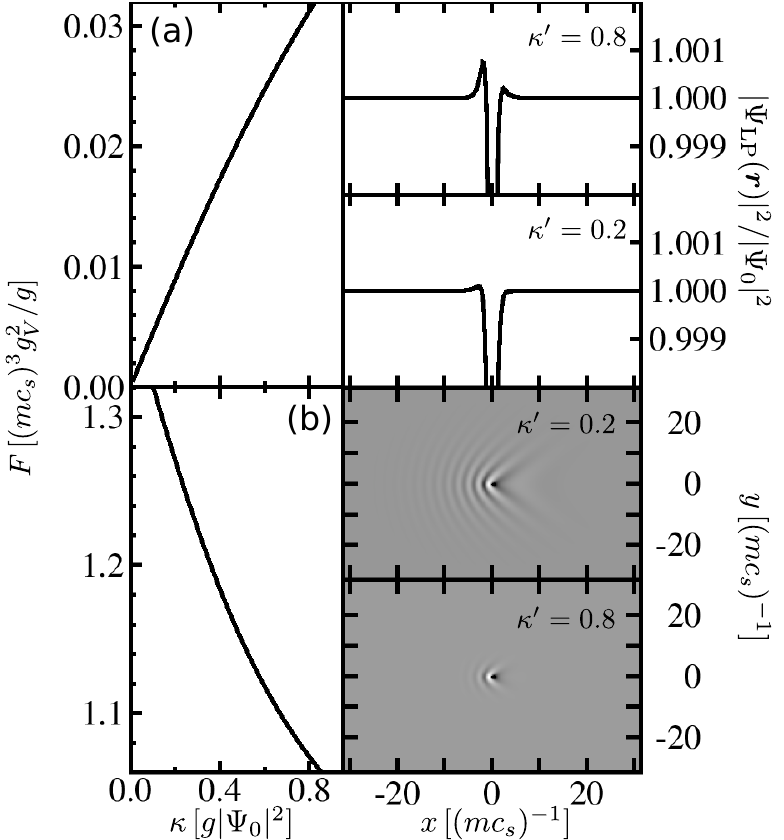}
\end{center}
\caption{(Color online) Drag force $F$ as a function of the inverse
  polariton lifetime $\kappa'=\kappa/(g |\psi_0|^2)$ in the (a)
  subcritical regime ($v_p=0.2 c_s$) and (b) supercritical regime
  ($v_p=1.9 c_s$). In both cases we have fixed
  $\Delta_p=-0.3g|\Psi_0|^2$ ($v^c \simeq 1.46 c_s$) but these results
  are qualitatively similar for any other value of the pump
  detuning. We plot in the insets the normalised real-space
  wavefunction $|\psi_{LP}(\vect{r})|^2/|\psi_0|^2$ for two specific
  values of $\kappa'=0.2$ and $\kappa'=0.8$.}
\label{fig:dragk}
\end{figure}
%
\section{Drag force}
\label{sec:drag}
The steady state response of the system to a static and weak defect
can be evaluated starting from Eq.~\eqref{eq:linre}:
\begin{equation*}
  \begin{pmatrix} \delta \psi_s(\vect{r}) \\ \delta
    \psi_s^*(\vect{r}) \end{pmatrix} =
  \hat{\mathcal{L}}^{-1} \begin{pmatrix} V(\vect{r}) e^{i \vect{k}_p
      \cdot \vect{r}} \psi_0 \\ -V(\vect{r}) e^{-i \vect{k}_p \cdot
      \vect{r}} \psi_0^{\star} \end{pmatrix} \; .
\end{equation*}
For a point-like defect, this can be written in momentum space as:
\begin{equation*}
  \delta \psi_s (\vect{k} + \vect{k}_p) = \Frac{-g_V \psi_0
    (\varepsilon(\vect{k}) - \vect{k} \cdot \vect{v}_p +
    i\kappa)}{\varepsilon(\vect{k}) [\varepsilon(\vect{k}) +
      2g|\psi_0|^2] - (\vect{k} \cdot \vect{v}_p - i\kappa)^2} \; ,
\end{equation*}
while the other component $\delta \psi_s^* (\vect{k}_p - \vect{k})$
can be obtained by complex conjugation and by substituting $\vect{k}
\mapsto -\vect{k}$. The drag force exerted by the defect on the fluid
is given by~\cite{astrakharchik04}:
\begin{equation}
  \vect{F} = - \int d\vect{r} |\psi_{LP}(\vect{r},t)|^2 \vect{\nabla}
  (V(\vect{r})) \; ,
\end{equation}
and, in the steady state linear response regime, we obtain:
\begin{multline}
  \vect{F} = g_V \int \frac{d\vect{k}}{(2\pi)^2} i\vect{k}
  \left[\psi_0^* \delta\psi_s (\vect{k} + \vect{k}_p) + \psi_0 \delta
    \psi_s^* (\vect{k}_p - \vect{k})\right]\\
  = 2g_V^2|\psi_0|^2 \int \frac{d\vect{k}}{(2\pi)^2} \frac{i\vect{k}
    \varepsilon(\vect{k})}{\omega^{+} (\vect{k})\omega^{-} (\vect{k})}
  \; .
    \label{eq:dragf}
\end{multline}
The drag is clearly oriented along the fluid velocity $\vect{v}_p$,
i.e., $\vect{F} = F \hat{\vect{v}}_p$. If $\kappa \to 0$, then the
integral in Eq.~\eqref{eq:dragf} is finite only if poles exist when
$\Re [\omega^{\pm} (\vect{k})] = 0$, i.e., when quasiparticles can be
excited, in agreement with the Landau criterion. For finite polariton
lifetimes, however, it is clear that the integral will be always
different from zero for $v_p>0$.
We now analyse the behaviour of the drag force as a function of the
fluid velocity for the three ($\Delta_p = 0$, $\Delta_P > 0$, and
$\Delta_p < 0$) different spectra illustrated in the previous section.

For the \emph{linear} spectrum, at $\Delta_p=0$, in the equilibrium
limit, $\kappa \to 0$, we recover for the drag the known result of
weakly interacting Bose gases in two
dimensions~\cite{astrakharchik04}:
\begin{equation}
  \frac{F}{(mc_s)^3 g_V^2/g}=\Frac{(v_p/c_s)^2 - 1}{v_p/c_s}
  \Theta(v_p - c_s)\; ,
\label{eq:drag0}
\end{equation}
with a threshold-like behaviour at a critical fluid velocity equal to
the speed of sound $c_s$. This limiting result is plotted as a bold
gray line in the panels (b,c) of Fig.~\ref{fig:dragv}. For
$\Delta_p=0$ and finite lifetimes $\kappa$, we find a smooth crossover
from the subsonic to the supersonic regime, with the drag being closer
to the equilibrium threshold behaviour for decreasing $\kappa$ (see
Fig.~\ref{fig:dragv}(b)). A finite lifetime tends to increase the
value of the drag in the subsonic region $v_p \ll v^c$, giving place
to a residual drag force, similar to what was found in the numerical
simulations of Ref.~\cite{cancellieri_10}. Instead, in the supersonic
region $v_p \gg v^c$, the finite lifetime tends to decrease the value
of the drag. In the case of \emph{diffusive-like} spectra at
$\Delta_p>0$ the situation is qualitatively very similar to the
resonant case (see Fig.~\ref{fig:dragv}(c)), with the difference that
now, in order to have stable solutions, we can decrease the value of
the lifetime only by decreasing accordingly also the value of the pump
detuning $\Delta_p$. The crossover for both $\Delta_p = 0$ and
$\Delta_p > 0$ is also qualitatively very similar to the case of
non-resonantly pumped polaritons~\cite{wouters10_b}, where the
spectrum of excitation is in that case diffusive-like.

In the case of \emph{gapped} spectra, the situation is however
qualitatively different (see Fig.~\ref{fig:dragv}(a)). For infinitely
living polaritons, $\kappa \to 0$, the drag force can also be
evaluated analytically and its expression is similar to
Eq.~\eqref{eq:drag0}, but with a critical velocity larger than the
speed of sound, which expression is given in Eq.~\eqref{eq:criti}:
\begin{equation}
  \frac{F}{(mc_s)^3 g_V^2/g}=\Frac{(v_p/c_s)^2 - 1}{v_p/c_s}
  \Theta(v_p - v^c)\; .
\end{equation}
Therefore now the drag experiences a jump for $v_p=v^c$, rather than a
continuous threshold as for the resonant case $\Delta_p=0$. As already
mentioned in the previous section, this discontinuous behaviour of the
drag for the gapped spectra is connected to the fact that, as soon as
quasiparticles can be excited by the defect at $v_p\ge v^c$,
Cherenkov-like waves can be immediately emitted in all directions,
rather than being restricted in a region outside the Mach cone like
before. For $\Delta_p=0$, the cone was gradually closing with
increasing the fluid velocity.

Both the increase of the value of the drag in the subcritical region
as a function of the polariton lifetime and the decrease in the
supercritical region, are behaviours common to all the types of
spectra. We plot the drag force as a function of $\kappa$ in
Fig.~\ref{fig:dragk}, for two values of the fluid velocity $v_p$ and a
specific value of the pump detuning $\Delta_p$, though we have checked
that the following results are generic. For $v_p < v^c$, we find that
the residual drag is a finite-lifetime effect only, and, in agreement
with the results of Ref.~\cite{cancellieri_10}, we find that, well
below the critical velocity, the drag force goes linearly to zero for
$\kappa \to 0$. In the resonant case $\Delta_p=0$, the slope of the
drag for $v_p \ll c_s$ can be evaluated analytically starting from the
expression~\eqref{eq:dragf}:
\begin{equation*}
    \Frac{F}{(mc_s)^3 g_V^2/g} \Simiq_{\kappa \to 0} \frac{2 c_s}{\pi
      v_p} \left( \frac{1}{ \sqrt{1-(v_p/c_s)^2}} - 1 \right)
    \Frac{\kappa}{g |\psi_0|^2} \; .
\end{equation*}
The residual drag in the subsonic regime is an effect of the
broadening of the quasi-particles energies: Even when the spectrum
real part does not allow any scattering against the defect (e.g., for
$\Delta_p \le 0$), the broadening produces some scattering close to
the defect. This results in a perturbation of the fluid around the
defect, asymmetric in the direction of the fluid velocity (see panel
(a) of Fig.~\ref{fig:dragk}), similarly to what it was obtained in
Ref.~\cite{cancellieri_10}. Instead, in the supersonic regime, the
drag force is weaker in the non-equilibrium case respect to the
equilibrium one. This is caused by the finite lifetime tending to
suppress the propagation of the Cherenkov waves away from the defect,
as shown in panel (b) of Fig.~\ref{fig:dragk}.

\section{Conclusions and discussion}
\label{sec:concl}
To conclude, we have analysed the linear response to a weak defect of
resonantly pumped polaritons in the pump-only state and we have been
able to determine two different kinds of threshold like behaviours for
the drag force as a function of the fluid velocity. In the case of
either zero or positive pump detuning, one can continuously connect to
the case of equilibrium weakly interacting gases, where the drag
displays a continuous threshold with a critical velocity equal to the
speed of sound. However, for negative pump detuning, where the
spectrum of excitations is gapped, the drag shows a discontinuity with
a critical velocity larger than the speed of sound. In this sense, the
case of coherently driven microcavity polaritons in the pump-only
configuration displays a richer phenomenology than the case of
polariton superfluids non resonantly pumped. It would be interesting
to perform a similar analysis in the case of polaritons in the optical
parametric oscillator regime, where polaritons are parametrically
scattered from the pump state to the signal and idler states. Here,
the spectrum of excitations has been already determined in
Ref.~\cite{wouters06b}, however it is far from clear what are the
conditions for subcritical, superfluid, behaviour in a fluid
characterised by three distinct currents, and how the link between
signal and idler imposed by the parametric scattering influences the
scattering of both fluids against a defect.

\subsection*{Acknowledgments}
 We are grateful to C. Tejedor and M. Szymanska for useful
 discussions. The authors acknowledge the financial support from the
 Spanish MINECO (MAT2011-22997), CAM (S-2009/ESP-1503), FP7 ITN
 "Clermont4" (A.B.), and from the program Ram\'on y Cajal (F.M.M.).
\\

\appendix
\section{Gross-Pitaevskii equation for the lower polariton field}
\label{app:full}
If one starts from a descriptions of polaritons in terms of separate
exciton and cavity photon fields, a rotation into the lower and upper
polariton basis, followed by neglecting the occupancy of the upper
polariton branch, results in the following Gross-Pitaevskii equation
for the lower polariton (LP) field in momentum space
$\psi_{LP}(\vect{r},t) = \sum_{\vect{k}} e^{i\vect{k}\cdot \vect{r}}
\psi_{LP,\vect{k}} (t)$~\cite{ciuti03}:
\begin{multline}
  i\partial_t \psi_{LP,\vect{k}} = f_p e^{-i\omega_p t}
  \delta_{\vect{k},\vect{k}_p} + \left[\omega_{LP} (k) - i\kappa
    (k)\right]\psi_{LP,\vect{k}} +\\
  \sum_{\vect{k}_1, \vect{k}_2} g_{\vect{k}, \vect{k}_1, \vect{k}_2}
  \psi^*_{LP,\vect{k}_1 + \vect{k}_2-\vect{k}} \psi_{LP,\vect{k}_1}
  \psi_{LP,\vect{k}_2} +\\ s_k \sum_{\vect{k}_1} V_{\vect{k} -
    \vect{k}_1} \psi_{LP,\vect{k}_1} s_{k_1}\; ,
\label{eq:efflp}
\end{multline}
where $\kappa(k)=\kappa_X c^2_k + \kappa_C s^2_k$ is the effective LP
decay rate, $g_{\vect{k}, \vect{k}_1, \vect{k}_2}=g_X c_{k}
c_{|\vect{k}_1 + \vect{k}_2-\vect{k}|} c_{k_1} c_{k_2}$ is the
interaction strength, and where $V(\vect{r}) = \sum_{\vect{k}}
e^{i\vect{k}\cdot \vect{r}} V_{\vect{k}}$. In these expressions, the
coefficients
\begin{equation}
  c^2_{k}, s^2_{k} = \Frac{1}{2} \left(1 \pm \Frac{\omega_C(k) -
    \omega_X^0}{\sqrt{(\omega_C(k) - \omega_X^0)^2 +
      \Omega_R^2}}\right)
\end{equation}
are the Hopfield coefficients used to diagonalise the free polariton
Hamiltonian. We want here to justify the simplified description done
in Eq.~\eqref{eq:basic}. If we follow the linear response expansion as
in~\eqref{eq:mfield}, the operator $\hat{\mathcal{L}}$ in momentum
space analogous to~\eqref{eq:opell} reads as:
\begin{widetext}
\begin{equation}
  \mathcal{L}_{\vect{k},\vect{k}_p} = \begin{pmatrix}
    \widetilde{\omega_{LP}} (\delta \vect{k}+\vect{k}_p) - i
    \kappa(\delta \vect{k}+\vect{k}_p) & g_X c_{k_p}^2 c_{\delta
      \vect{k}+\vect{k}_p} c_{\delta \vect{k}-\vect{k}_p} \psi_0^2
    \\ - g_X c_{k_p}^2 c_{\delta \vect{k}+\vect{k}_p} c_{\delta
      \vect{k}-\vect{k}_p}{\psi_0^{\star}}^2 & -
    \widetilde{\omega_{LP}}(\delta \vect{k}-\vect{k}_p) -
    i\kappa(\delta \vect{k}-\vect{k}_p) \end{pmatrix}\; ,
\label{eq:opel2}
\end{equation}
\end{widetext}
where now $\widetilde{\omega_{LP}} (\delta \vect{k} \pm\vect{k}_p) =
\omega_{LP} (\delta \vect{k} \pm\vect{k}_p) -\omega_p + 2 g_X
c_{k_p}^2 c_{\delta \vect{k} \pm \vect{k}_p}^2 |\psi_0|^2$. It is easy
to show that the eigenvalues of this operator coincide with our
approximated expressions~\eqref{eq:spect} in the limit of $\delta k
\ll k_p$, when $c_{\delta \vect{k} \pm \vect{k}_p}^2 \simeq
c_{k_p}^2$, $s_{\delta \vect{k} \pm \vect{k}_p}^2 \simeq s_{k_p}^2$
and when we can simply rename $g=g_X c_{k_p}^4$ and $\kappa =
\kappa(k_p)$. It is interesting to note that, even if we would retain
the linear terms in $\vect{k}_p \cdot \delta \vect{k}$ in the
expansion of $c_{\delta \vect{k} \pm \vect{k}_p}^2$, this would result
in a renormalisation of the fluid velocity $\vect{v}_p$ in the
expression~\eqref{eq:spect} which takes into account the blue-shift of
the lower polariton dispersion due to the interaction.


\newcommand\textdot{\.}

\end{document}